\documentclass{article}

\usepackage{spconf,amsmath,graphicx,hyperref}
\usepackage{microtype}
\usepackage{amssymb}
\usepackage{booktabs}
\usepackage{makecell}
\usepackage{colortbl}
\usepackage{xcolor}
\usepackage[ruled,linesnumbered,lined]{algorithm2e}

\title{FP-ANet: A fixed-point Attention Network for Hybrid-field THz Ultra-Massive MIMO Channel Estimation}
\name{Kangchun Zhao$^{\dagger}$, Haitian Yang$^{\dagger}$, Yijie Mao$^{\star}$\thanks{ $^{\dagger}$Co-first author, $^{\star}$ Corresponding author }
\thanks{This work has been supported by the National Nature Science Foundation of China under Grant 62571331.}}
\address{School of Information Science and Technology, ShanghaiTech University, Shanghai, China}
\begin{document}
\ninept
\maketitle
\begin{abstract}
Ultra-massive multiple-input multiple-output (UM-MIMO) is a key technology for enabling terahertz (THz) communications in 6G networks, offering high beamforming gain to combat severe path loss. 
However, the large antenna array expands the near-field region, resulting in a hybrid near- and far-field communication environment. 
This makes channel estimation significantly more challenging than in conventional networks.
To address this issue, we propose a novel attention augmented channel estimator named the fixed-point attention network~(FP-ANet), which integrates fixed-point theory with a dual-attention mechanism. By combining a linear and dual-attention residual blocks based non-linear estimator in each iteration, this model-driven approach effectively exploits the sparsity of THz channels in the angular-distance domain, enabling a more precise and physically-grounded channel estimation. Simulation results show that FP-ANet achieves superior channel estimation accuracy compared to state-of-the-art methods while maintaining comparable computational complexity.
\end{abstract}
\section{Introduction}
\label{sec:intro}
The terahertz (THz) band is a key enabling technology for future wireless systems~\cite{han2024cross, zeng2025hybrid}, offering vast bandwidth to support ultra-high data rates. However, THz signals suffer from severe propagation attenuation due to spreading loss and molecular absorption. Fortunately, the sub-millimeter wavelength at THz frequencies allows dense integration of antennas into compact arrays, enabling ultra-massive multiple-input multiple-output (UM-MIMO) systems that provide substantial beamforming gain to overcome these coverage limitations~\cite{han2024cross}.
To realize the full potential of THz UM-MIMO, accurate channel estimation is essential.
However, the large number of antennas significantly increases the array aperture, expanding the near-field region and resulting in a hybrid near- and far-field communication environment~\cite{zeng2025hybrid,tarboush2024cross, fang2025optimal}. 
This makes high-dimensional channel estimation considerably more challenging than in conventional uniform-field scenarios.


Several paradigms have been explored to address this complex estimation task. Conventional channel estimation approaches in THz UM-MIMO often treat the hybrid-field channel as a uni-field channel, leveraging either near-field or far-field properties with compressive sensing techniques to reduce pilot overhead~\cite{choi2017compressed,dovelos2021channel,chen2023hybrid}.
However, the estimation accuracy is severely degraded by the channel model mismatch. 
To enhance performance, advanced deep unfolding techniques are employed in \cite{hu2022ddpg,yu2023adaptive,hu2023prince}. These model-based deep learning methods learn channel characteristics by mapping the steps of an iterative algorithm onto a fixed number of neural network layers.
However, such an approach cannot guarantee convergence and may degrade the performance~\cite{qiao2025adaptive, sibio2024low,lin2025ris}.
In contrast, frameworks based on the fixed-point theory can guarantee both the convergence to a unique fixed point and adaptive complexity of the estimator by applying learned parameters to each iteration.
For instance, frameworks like  fixed point network orthogonal approximate message passing (FPN-OAMP)~\cite{yu2023adaptive} leverage the fixed-point theory to guarantee convergence. However, their practical performance is ultimately constrained by a non-linear estimator (NLE) that provides an isotropic treatment to all channel features. This inherent limitation prevents it from effectively exploiting the known sparse structure of the THz channel, leading to a significant performance bottleneck.

To further improve channel estimation accuracy, we propose a novel fixed-point attention network (FP-ANet), a channel estimator that integrates fixed-point theory with a dual-attention mechanism.
The proposed fixed-point network contains two main parts:
a linear estimator (LE) module and a non-linear estimator (NLE) module.
We propose a dual-attention NLE model that leverages both channel and spatial attention mechanisms to exploit the sparsity of the THz channel, identifying and amplifying the most salient features across channel and spatial dimensions.
This enables the estimator to intelligently focus on sparse components in the angular-distance domain, thereby achieving superior accuracy. 
Furthermore, by building upon fixed-point theory, the convergence of our FP-ANet is rigorously guaranteed. Simulation results demonstrate that FP-ANet outperforms state-of-the-art methods in estimation accuracy with comparable computational complexity.

\section{System Model and Problem Formulation}
\label{sec:format}
\subsection{System Model}

We consider an uplink THz UM-MIMO network where multiple single-antenna user equipments (UEs) transmit pilot signals to a base station (BS) with a planar array-of-subarrays (AoSA) architecture. 
The AoSA is a uniform planar array situated entirely in the $x-y$ plane, composed of $N$ identical subarrays (SAs) arranged in a square grid of $\sqrt{N}\times \sqrt{N}$.
Each SA is a uniform planar array containing $\bar{N}$
antenna elements (AEs) arranged in a $\sqrt{\bar{N}}\times \sqrt{\bar{N}}$ grid. The BS is equipped with  a total of $M =N \bar N$ AEs.
The index $n$ of an SA at the $x$-th row and $y$-th column of the main AoSA grid is given by $n = (x-1)\sqrt{N}+y$, where $1\leq x,y\leq \sqrt{N}$.
For a given SA, the index $\bar{n}$ of an AE at the $\bar{x}$-th row and $\bar{y}$-th column within the given SA is defined by $\bar{n} =(\bar{x}-1)\sqrt{\bar{N}}+\bar{y}$, where $1\leq \bar{x},\bar{y}\leq \sqrt{\bar{N}}$.
The spacing between adjacent AEs is denoted by $d_{\text{a}}$, while the spacing between adjacent SAs is $d_{\text{sub}}$.
We establish a Cartesian coordinate system with its origin at the center of the first AE in the first SA. 
With this configuration, the coordinate vector $\mathbf{p}_{n,\bar{n}}$ for the $\bar{n}$-th AE within the $n$-th SA is given by:
\begin{equation}
    \mathbf{p}_{n,\bar{n}} = 
\begin{pmatrix} 
(x-1)\left[(\sqrt{\bar{N}}-1)d_{\text{a}}+d_{\text{sub}}\right]  + (\bar{x}-1)d_{\text{a}}  \\ 
(y-1)\left[(\sqrt{\bar{N}}-1)d_{\text{a}}+d_{\text{sub}}\right] + (\bar{y}-1)d_{\text{a}} \\ 
0 
\end{pmatrix}.
\end{equation}
The practical partially-connected
hybrid beamforming architecture is applied at the BS, where all AEs in each SA share the same radio frequency (RF) chain, resulting in a total of $N$ RF chains at the BS~\cite{zhang2019hybrid}.
The propagation environment encompasses a hybrid of far-field and near-field effects. The boundary between these two regions is determined by the Rayleigh distance $D_{\text{Rayleigh}} = \frac{2D^2}{\lambda_c}$, where $D$ is the array aperture and  $\lambda_c$
 is the carrier wavelength.
The channels between the BS and UEs exhibit a mixture of near-field and far-field features due to large AoSA aperture and short THz wavelength, with the number of each path type being variable.
Considering the property of THz waves, there is limited scattering. The spatial-frequency channel $\tilde{\mathbf{h}}\in\mathbb{C}^{M\times 1}$ between the BS and an UE can be described using a combination of a single LoS path and $L-1$ non-LoS paths~\cite{dovelos2021channel,yao2023blind}, modeled as
\begin{align}
    \tilde{\mathbf{h}} = \sum_{l=1}^{L} \beta_l \mathbf{a}(\phi_l, \theta_l, r_l)e^{-j2\pi f_c\tau_l},
\end{align}
where $f_c$ is the carrier frequency,  $\beta_l$ is the complex path gain, while $\phi_l,
\theta_l, r_l, \tau_l$ represent the azimuth angle, elevation angle, distance, time delay of the $l$-th path, respectively. 
$\mathbf{a}(\phi_l,
\theta_l, r_l) \in \mathbb{C}^{M\times 1}$ denotes the array response vector, and its form differs for near-field and far-field paths depending on the distance $r_l$.
\begin{align}
     \mathbf{a}(\phi_l, \theta_l, r_l) = 
\begin{cases} 
\mathbf{a}^{\text{near}}(\phi_l, \theta_l, r_l), & \text{if } r_l < D_{\text{Rayleigh}}, \\
\mathbf{a}^{\text{far}}(\phi_l, \theta_l, r_l), & \text{otherwise}.
\end{cases}
\end{align}
The array response vectors $\mathbf{a}^{\text{near}}(\phi_l, \theta_l, r_l)$ and $\mathbf{a}^{\text{far}}(\phi_l, \theta_l, r_l)$ for the near and far fields are obtained as follows. 
Given the spherical wavefront in the near-field, every element of  the array response matrix depends on the precise distance from each AE to the RF source/scatterer.
For the $l$-th path, the location of the RF scatter is identified by the vector $r_l\mathbf{x}_l$, where $\mathbf{x}_l$ is a unit vector that indicates the angle of arrival (AoA) direction, defined by $\mathbf{x}_l = (\sin(\theta_l)\cos(\phi_l),\sin(\theta_l)\sin(\phi_l),\cos(\theta_l))^T$.
Based on this, the response matrix for the $\bar{n}$-th AE located within the $n$-th SA is 
\begin{equation}
    \left( \mathbf{A}^{\text{near}}(\phi_l, \theta_l, r_l) \right)_{n,\bar{n}} = e^{-j2\pi \frac{\left\| \mathbf{p}_{n,\bar{n}} - r_l \mathbf{x}_l \right\|_2}{\lambda_c} }
\end{equation}
Through vectorization of  $\mathbf{A}^{\text{near}}(\phi_l, \theta_l, r_l)$, we obtain the near-field array response vector as $\mathbf{a}^{\text{near}}(\phi_l, \theta_l, r_l) = \text{vec}(\mathbf{A}^{\text{near}}(\phi_l, \theta_l, r_l))$.
Given the planar wavefront in the far-field, the response matrix is 
\begin{equation}
    \left( \mathbf{A}^{\text{far}}(\phi_l, \theta_l, r_l) \right)_{n,\bar{n}} = e^{-j2\pi \frac{\mathbf{p}^T_{n,\bar{n}}\mathbf{x}_l}{\lambda_c}}.
\end{equation}
The far-field  array response vector is obtained as 
$\mathbf{a}^{\text{far}}(\phi_l, \theta_l, r_l) = \text{vec}(\mathbf{A}^{\text{far}}(\phi_l, \theta_l, r_l))$.

\subsection{Problem formulation}
To estimate the uplink channel between the BS and UE, UEs transmit pilot signals to the BS  across $T$ time slots.
Assuming orthogonal pilots, we analyze an arbitrary UE without loss of generality.
For the ease of algorithm design, we transform $\tilde{\mathbf{h}}$ to its  angular domain by using $\bar{\mathbf{h}}=\mathbf{E}\tilde{\mathbf{h}}$, where   $\mathbf{E}=\text{blkdiag}(\mathbf{U}_1,\ldots,\mathbf{U}_N)$  is a unitary matrix
 with $\mathbf{U}_n,n\in\{1,\ldots,N\}$ being a $\bar{N}\times \bar{N}$ discrete Fourier transform (DFT) matrix.
The received signal $\mathbf{y}_t \in \mathbb{C}^{N \times 1}$  observed at the $t$-th time slot is given as
\begin{equation}
   \mathbf{y}_t = \mathbf{D}_{t}^H \mathbf{W}_t^H(\mathbf{E}\bar{\mathbf{h}}s_t +\mathbf{n}_t)
= \mathbf{D}_{t}^H \mathbf{W}_t^H \mathbf{E}\bar{\mathbf{h}}s_t + \mathbf{D}_{t}^H \mathbf{W}_t^H \mathbf{n}_t,
\end{equation}
 where $\mathbf{D}_{t}\in \mathbb{C}^{N \times N}$ is the digital combining matrix and $\mathbf{W}_{t}\in \mathbb{C}^{N\bar{N} \times N}$ equal to $\mathrm{blkdiag}(\mathbf{w}_{1,t}, \mathbf{w}_{2,t}, \dots, \mathbf{w}_{N,t})$ is the analog combining matrix. $\mathbf{w}_{i,t} \in \mathbb{C}^{\bar{N} \times 1}$ is subject to a constant-modulus constraint.
 The pilot symbol $s_t$, known at the BS, is set to 1 for convenience. $\mathbf{n}_t \sim \mathcal{CN}(\mathbf{0}, \sigma_n^2 \mathbf{I})$ denotes the additive white Gaussian noise (AWGN).
An optimal configuration of the combining matrices is not feasible without prior channel knowledge. 
We therefore set the digital matrix $\mathbf{D}_{t}$  to be the identity matrix $\mathbf{I}$ for an arbitrary scenario.
Meanwhile, to save energy consumption, the analog phase shifts within $\mathbf{W}_{t}$ are chosen randomly from a one-bit quantized set, i.e., $(\mathbf{w}_{i,t})_j \in \frac{1}{\sqrt{\bar{N}}}\{\pm 1\}$~\cite{he2018deep}.
The  received pilot signal, after aggregating all $T$ time slots is  $\bar{\mathbf{y}} = [\mathbf{y}_1^T, \mathbf{y}_2^T, \dots, \mathbf{y}_T^T]^T \in \mathbb{C}^{NT \times 1}$ and can be expressed by
\begin{equation}
    \bar{\mathbf{y}}=\bar{\mathbf{M}} \bar{\mathbf{h}}+\bar{\mathbf{n}},
\end{equation}
 where $\bar{\mathbf{M}}=[(\mathbf{W}_{1}^H \mathbf{E})^T, \dots, (\mathbf{W}_{T}^H \mathbf{E})^T]^T\in \mathbb{C}^{NT \times N\bar{N}}$, $\bar{\mathbf{n}}  =[(\mathbf{W}_{1}^H \mathbf{n}_1)^T, \dots, (\mathbf{W}_{T}^H \mathbf{n}_T)^T]^T \in \mathbb{C}^{NT \times 1}$.
To transform $\bar{\mathbf{y}}$ into an equivalent real-valued representation, we map the  complex vectors  to new real vectors combining 
their real and imaginary components, denoted as $\mathbf{y} = [\Re(\bar{\mathbf{y}})^T, \Im(\bar{\mathbf{y}})^T]^T \in \mathbb{R}^{2NT \times 1}$, $\mathbf{h} = [\Re(\bar{\mathbf{h}})^T, \Im(\bar{\mathbf{h}})^T]^T \in \mathbb{R}^{2N\bar{N} \times 1}$,  $\mathbf{n} = [\Re(\bar{\mathbf{n}})^T, \Im(\bar{\mathbf{n}})^T]^T \in \mathbb{R}^{2NT \times 1}$ and 
\begin{equation}
    \mathbf{M} = \begin{pmatrix} \Re(\bar{\mathbf{M}}) & -\Im(\bar{\mathbf{M}}) \\ \Im(\bar{\mathbf{M}}) & \Re(\bar{\mathbf{M}}) \end{pmatrix}\in\mathbb{R}^{2NT\times 2N\bar{N}}. 
\end{equation}
Then, the received signal after $T$ time slots is  transformed to  
\begin{equation}
    \mathbf{y} = \mathbf{M}\mathbf{h} + \mathbf{n}.
\end{equation}
Our goal is to estimate the original channel $\mathbf{h}$ from the received signal $\mathbf{y}$ with noise $\mathbf{n}$ and the  measurement matrix $\mathbf{M}$.
 Due to the existence of both far-field and near-field components in the channel, it is difficult to obtain an accurate estimation using the existing uni-field methods.
To overcome this bottleneck, 
we propose the FP-ANet to estimate the channel in the following.


\section{Proposed fixed-point  attention-based network (FP-ANet)}
\label{algorithm}
Conventional iterative channel estimation methods such as OAMP algorithm can be mathematically framed as a search for a fixed point of a specific mapping function. 
Motivated by the fixed-point theory, for the model-driven DL framework, each iteration  can be represented by the general form $\mathbf{h}^{(k+1)} = f_{\Phi}(\mathbf{h}^{(k)}; \mathbf{y})$, where $\mathbf{h}^{(k)}$ denotes the estimated channel at the $k$-th iteration and $f_{\Phi}(\cdot)$ is the mapping function parameterized by $\Phi$.
$\Phi$ denotes the complete set of learnable parameters of the neural network embedded within the mapping function.
Assuming the iterative sequence converges, its limit as $k\to \infty$,  denoted as $\mathbf{h}^{*}$, should satisfy the equation
\begin{equation}
    \mathbf{h}^{*} = f_{\Phi}(\mathbf{h}^{*}; \mathbf{y}).
\end{equation}
This equation mathematically describes the state of equilibrium for the algorithm, modeling its behavior precisely at the point of convergence.
The key challenge is to design the mapping function $f_{\Phi}$ such that its convergence to a unique fixed point is guaranteed.
Leveraging Banach fixed-point theorem in \cite{bauschke2020correction}, if $f_{\Phi}(\cdot; \mathbf{y})$
is a contraction mapping with Lipschitz constant $L$, the sequence $\{\mathbf{h}^{(k)}\}$ updated through $\mathbf{h}^{(k+1)} = f_{\Phi}(\mathbf{h}^{(k)}; \mathbf{y})$ converges to the unique fixed point $\mathbf{h}^*$ at a linear rate, guaranteeing that the designed DL-based mapping function $f_{\Phi}(\cdot; \mathbf{y})$ converges to a fixed point $\mathbf{h}^{*}$.
In this paper, we propose the FP-ANet, a novel deep learning architecture for hybrid-field THz channel estimation. FP-ANet is built upon the fixed-point framework and attention mechanism, which iteratively refines a channel estimation using two modules: a LE module and an attention-based NLE module.
\begin{figure}[t!]
    \centering
\includegraphics[width=0.4\textwidth]{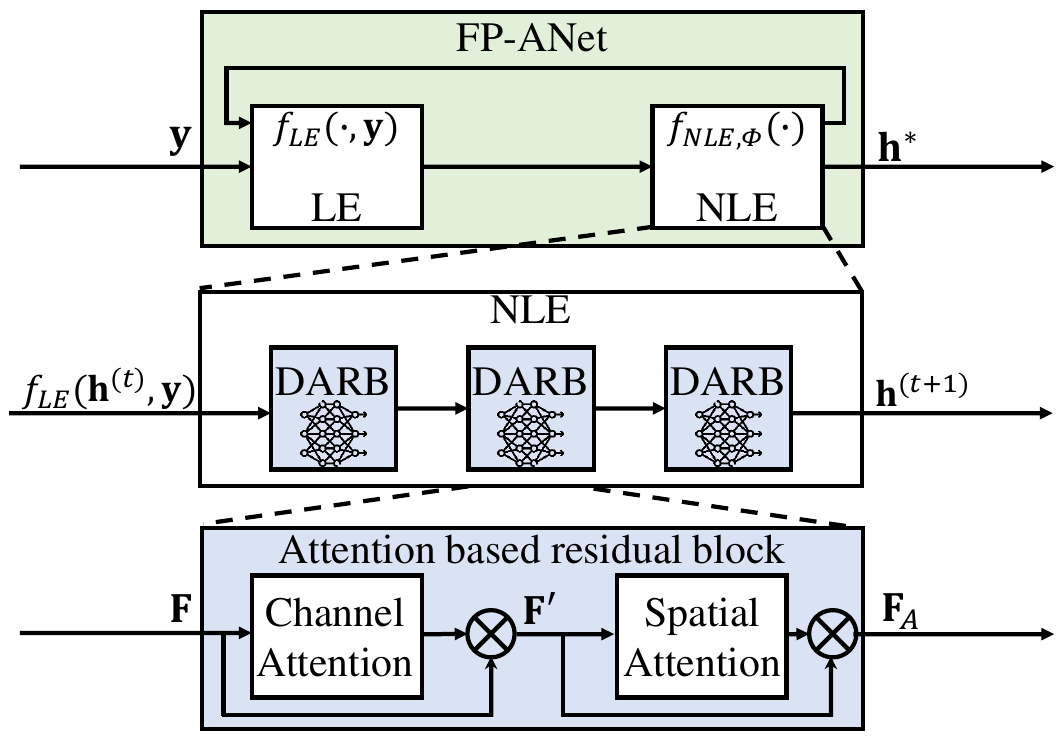}
    \caption{Network architecture of the FP-ANet.}
    \label{fig:fig1}
\end{figure}
The mapping $f_{\Phi}$  represents the combined operation of these two modules, i.e., $ f_{\Phi}(\mathbf{h}; \mathbf{y}) = (f_{\text{NLE}, \Phi} \circ f_{\text{LE}})(\mathbf{h}; \mathbf{y})$.
The detailed framework of the FP-ANet are represented in Fig.~\ref{fig:fig1} and Algorithm~\ref{FPANet}.

\subsection{LE module}
Given the received pilot signal $\mathbf y$, we first obtain a coarse estimation of $\mathbf {h}$ by a linear function $f_{\text{LE}}(\cdot; \mathbf{y})$, which is denoted as 
\begin{equation}
    f_{	\text{LE}}(\mathbf{h}^{(k)}; \mathbf{y}) = \mathbf{h}^{(k)} + \mathbf{Z}(\mathbf{y} - \mathbf{M}\mathbf{h}^{(k)}).
\end{equation}
Leveraging the LE module in OAMP algorithm, $\mathbf{Z}\in\mathbb{R}^{2N\bar{N}\times 2NT}$ can be constructed as $\mathbf{Z} = \rho \mathbf{M}^\dagger = \frac{2N\bar{N}}{\mathrm{tr}(\mathbf{M}^\dagger \mathbf{M})} \mathbf{M}^\dagger,$
where $\mathbf{M}^\dagger$ denotes the pseudo inverse matrix of $\mathbf{M}$ and  the step size $\rho = \frac{2N\overline{N}}{\mathrm{tr}(\mathbf M^{\dagger}\mathbf M)}$ is rigorously derived from the OAMP framework~\cite{ma2017orthogonal}, satisfying the condition $\mathrm{tr}(\mathbf I - \rho \mathbf M^{\dagger}\mathbf M)=0$. This ensures that the output error of the LE module is uncorrelated with the input error, which is crucial for the optimality of the subsequent iterative estimation.
\subsection{NLE module}

The attention-based NLE is given as 
\begin{equation}
    f_{\text{NLE}, \Phi}\left(f_{\text{LE}}(\mathbf{h}^{(k)}; \mathbf{y}) \right)= A_{ \Phi} \left(f_{\text{LE}}(\mathbf{h}^{(k)}; \mathbf{y}) \right),
\end{equation}
where $A_{ \Phi}$ is a dual-attention structure containing three  dual-attention residual blocks (DARBs), depicted in Fig.~\ref{fig:fig1}. 
The  data pipeline of the NLE module  begins by reshaping the input vector $f_{\text{LE}}(\mathbf{h}^{(k)}; \mathbf{y})$ into a 2D feature map and projecting it into a high-dimensional space via an input convolution. 
  Next, three DARBs  refine the features before an output convolution maps them back to the dimension of the original channel, which is then flattened to output the final estimate $\mathbf{h}^{(k+1)}$.
In each DARB, there are two sequentially connected  attention blocks, as shown in Fig. \ref{fig:fig1}. This architecture is designed to exploit the inherent sparsity of the  THz channel by integrating the channel and spatial attention blocks. 

\textit{1) Channel Attention Block:}
The channel attention block performs feature recalibration to suppress noise artifacts by  adaptively weighting and amplifying  important channels of the input feature maps $\mathbf{F}\in\mathbb{R}^{H\times W\times C}$, where $H$, $W$ and $C$ denote the height, width, and the channel dimension of the feature map. 
$\mathbf{F}$ is  fed through a shared multi-layer perceptron (MLP) to generate
a channel attention map $\mathbf{F}_c\in\mathbb{R}^{1\times 1\times C}$.
The output of the channel attention block is then computed by 
\begin{equation}
    \mathbf{F}' = \mathbf{F}_c \otimes 
\mathbf{F},
\end{equation}
where $\otimes$ denotes element-wise multiplication.

\textit{2) Spatial Attention Block:}
After the channel attention block, the spatial attention block explicitly targets the angular-distance sparsity of the hybrid near-field by  generating a 2D mask $\mathbf{F}_s\in\mathbb{R}^{H\times W\times 1}$ that pinpoints the precise locations of those amplified components in the angular-distance domain.
The output of the spatial attention block is obtained by 
\begin{equation}
    \mathbf{F}_{A} = \mathbf{F}_s \otimes \mathbf{F}'.
\end{equation}
By designing the NLE module to  exploit the channel's inherent sparsity, its synergistic ``what and where'' refinement process can intelligently pinpoint and preserve sparse paths while suppressing noise, achieving highly accurate and structure-aware channel estimation.
\begin{algorithm}[t!]	
 	 \textbf{Input}:Received signal $\mathbf{y}$, measurement matrix $\mathbf{M}$, trained NLE parameters $\Phi$, error tolerance $\epsilon$\;
\textbf{Output}: The estimated THz channel $\hat{\mathbf{h}}$\;
 \textbf{Initialize}: $\mathbf{h}^{(0)} \gets \mathbf{0}$, $k \gets 0$\;
 	\Repeat{$\|\mathbf{h}^{(k)} - \mathbf{h}^{(k-1)}\|_2 < \epsilon$} 
 	{$k \gets k + 1$\;
 	  $\mathbf{h}^{(k)} \gets (f_{\text{NLE},\Phi} \circ f_{\text{LE}})(\mathbf{h}^{(k-1)}; \mathbf{y})$\;     
      }
       $\hat{\mathbf{h}}\gets \mathbf{h}^{(k)}$\;
 \caption{FP-ANet for THz hybrid channel estimation}
\label{FPANet}		
 \end{algorithm} 
\subsection{Convergence analysis}

Our proposed FP-ANet achieves linear convergence to a fixed point $\mathbf{h}^{*}$. 
To prove this, we need to show that $(f_{\text{NLE}, \Phi} \circ f_{\text{LE}})(\mathbf{h}; \mathbf{y})$  is a contractive mapping.
As proven in [Theorem 7]~\cite{yu2023adaptive}, the mapping $f_{\text{LE}}(\mathbf{h}^{(k)}; \mathbf{y})$ is Lipschitz continuous with Lipschitz constant equal to 1.
Furthermore, as established in~\cite{yu2023adaptive}, if the NLE module $f_{\text{NLE}, \Phi}$ is also Lipschitz continuous, then the composite mapping $f_{\Phi}$ is guaranteed to be Lipschitz continuous. To ensure $f_{\Phi}$ is a contraction, its Lipschitz constant must be less than 1.
To ensure that the overall mapping is a contraction, we first approximate the Lipschitz constant $f_{\text{NLE}, \Phi}(\cdot)$ after each iteration using equation (22) from \cite{yu2023adaptive}. If $\hat{L}>1$, we enforce the contraction property by normalizing the parameters $\Phi$, for instance, by applying the normalization method proposed in~\cite{yu2022hybrid}.

\subsection{Complexity analysis}
The computational complexity of our proposed FP-ANet mainly stems from two parts: the LE module and NLE module.
For the LE module, the complexity is primarily determined by  matrix-vector product, resulting in a computational cost of $\mathcal{O}(4N^2\bar{N}T)$.
For the NLE module, the complexity is determined by the number of floating-point operations (FLOPs) within its neural network architecture, denoted by a constant $s$. 
Based on fixed-point theory, FP-ANet exhibits linear convergence. This implies that the number of iterations required to achieve an $\epsilon$-optimal solution is $\mathcal{O}(\log\frac{1}{\epsilon})$.
The total computational complexity of the FP-ANet is given by $\mathcal{O}(\log\frac{1}{\epsilon}(4N^2\bar{N}T+s))$.
\begin{figure}[t]
    \centering
\includegraphics[width=0.38\textwidth]{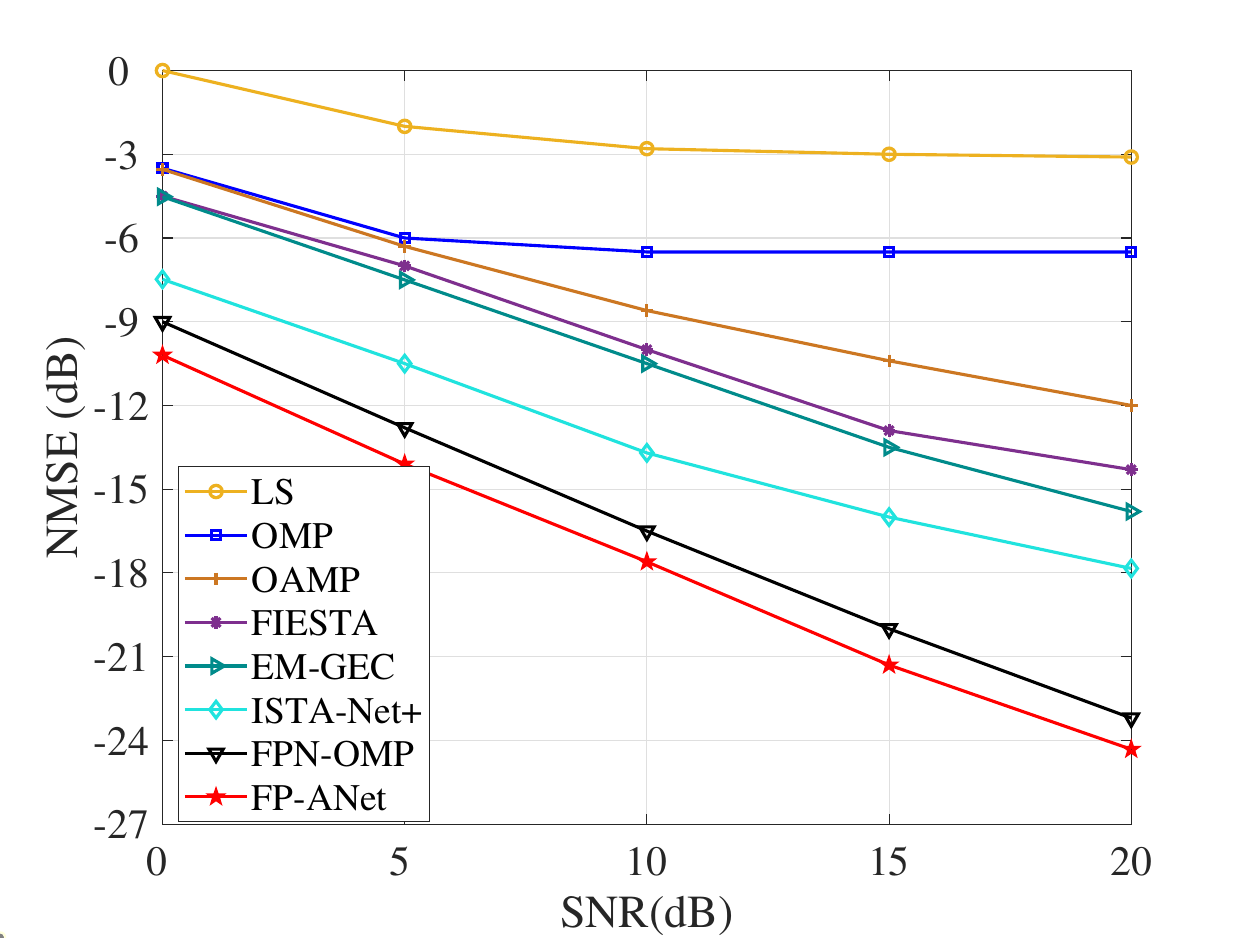}
    \caption{NMSE of channel estimation with various SNR levels.}
    \label{fig:fig2}
\end{figure}
\section{SIMULATION RESULTS}
In this section, we evaluate the performance of FP-ANet through  simulation results.

\subsection{Comparison with other baselines}
We compare its channel estimation accuracy with the  FPN-OAMP  and other state-of-the-art methods.
To ensure a fair and direct comparison, our simulation setup  follows the configuration described in~\cite{yu2023adaptive}. 
We consider an uplink THz UM-MIMO system where the BS is equipped with a planar AoSA comprising 1024 antennas and 4 RF chains.
The hybrid-field channel consists of  $L=5$ paths, with distances and angles configured to represent a realistic mixture of near-field and far-field components.
The training, validation, and testing datasets contain 80000, 5000, and 5000 samples, respectively.
We train the FP-ANet with batch size of 128 and convergence tolerance $\epsilon$ equal to 0.001.  
The primary metric for evaluation is the NMSE, defined as $\text{NMSE} = \mathbb{E}\left[ \frac{\|\mathbf{h} - \hat{\mathbf{h}}\|_2^2}{\|\mathbf{h}\|_2^2} \right]$.
We compare our FP-ANet against several  baselines including the least squares (LS), OMP~\cite{cui2022channel}, OAMP~\cite{dovelos2021channel}, FIESTA~\cite{beck2009fast}, EM-GEC~\cite{wang2019channel}, ISTA-Net+~\cite{jin2021adaptive} and FPN-OAMP~\cite{yu2023adaptive}.

\begin{figure}[t]
    \centering
\includegraphics[width=0.38\textwidth]{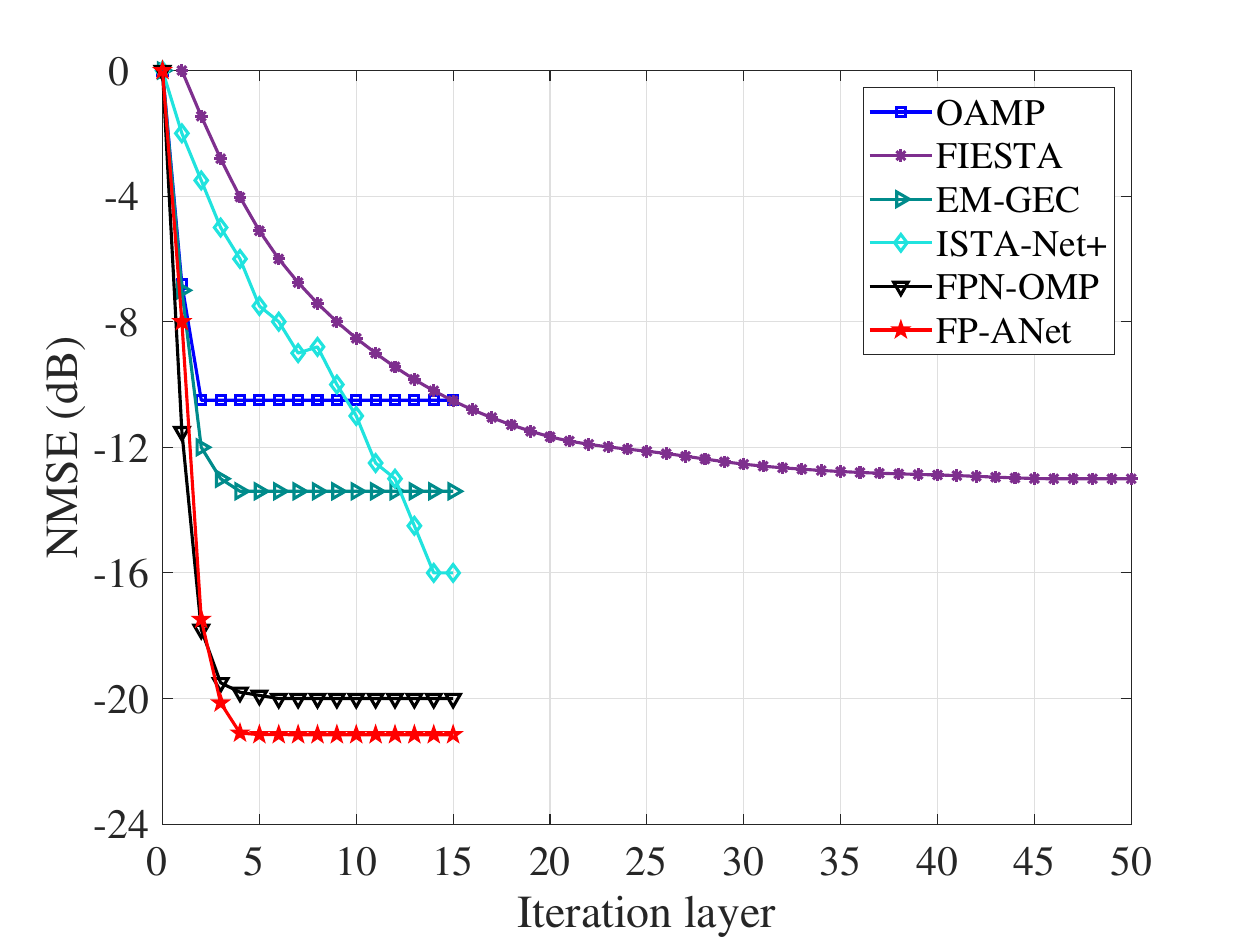}
    \caption{NMSE  versus the number of iterations at an SNR of 15 dB.}
    \label{fig:fig3}
\end{figure}

Fig.~\ref{fig:fig2} compares the NMSE performance of FP-ANet against the baseline methods across various SNRs. The results clearly show that FP-ANet consistently outperforms all baselines.
Notably, compared to the FPN-OAMP framework, the FP-ANet achieves a significant and stable NMSE reduction of approximately 1.5 dB across the entire tested SNR range.
This performance gain is attributed to the dual-attention module, which enables the NLE to effectively exploit the inherent channel sparsity. This capability is particularly advantageous at low SNRs, where the attention mechanism can better distinguish weak channel paths from strong noise, a task challenging for conventional architectures.
At low SNR, the attention mechanism demonstrates a superior ability to distinguish weak channel paths from strong noise. The convergence behavior, depicted in Fig.~\ref{fig:fig3}, further highlights the efficiency of our method. FP-ANet not only reaches a substantially lower NMSE floor but also converges rapidly, achieving near-optimal performance within just a few iterations.

\subsection{Ablation Study on the NLE Architecture}
We conduct an ablation study to verify the efficacy of our proposed dual-attention module within the NLE module, with the results detailed in Table~\ref{ablation_study}. For a fair comparison, the linear estimator remains unchanged across all variants; thus, we report the FLOPs associated solely with the NLE module. Replacing our module with a standard self-attention~\cite{vaswani2017attention} (SA) block of comparable complexity (see FP-ANet (SA-Light) in Table~\ref{ablation_study}) results in a significant NMSE degradation to -10.7 dB. This underscores that the performance gain stems from our structure-aware design, not merely the inclusion of an attention mechanism. Conversely, employing a larger SA block (see FP-ANet (SA-Large) in Table~\ref{ablation_study}) yields a marginal NMSE improvement but at a prohibitive computational cost, with a 42\% increase in FLOPs and more than double the runtime. In contrast, our proposed FP-ANet achieves a superior balance between performance and efficiency, attaining a -14.1 dB NMSE with significantly lower computational overhead.

\begin{table}[htbp]
\caption{Performance and efficiency analysis at an SNR of 5 dB. Time is measured as the per-epoch training duration on a single NVIDIA A40 GPU. FLOPs correspond to the non-linear estimator only.}
    \centering
    \begin{tabular}{lcccc} 
   \toprule
       \textbf{Method}  & \textbf{NMSE} & \textbf{Time}& \textbf{Params} & \textbf{FLOPs}  \\
    \midrule
      FPN-OAMP & -12.8dB & 72.7s & 362.3k & 59.3M\\
      FP-ANet (SA-Light) & -10.7dB & 120.4s & 256.6k & 59.8M\\
      FP-ANet (SA-Large) & -14.4dB & 128.0s & 364.9k & 84.9M \\
      \rowcolor{blue!5}FP-ANet & -14.1dB & 64.4s & 364.9k & 59.8M\\
    \bottomrule
    \end{tabular}
    
    \label{ablation_study}
    
\end{table}

\section{CONCLUSION}
In this paper, we propose an attention-based deep learning estimator FP-ANet for  hybrid channel estimation in the THz  UM-MIMO system. 
By replacing the conventional ResNet-based module with a dual-attention mechanism as the non-linear estimator, FP-ANet effectively leverages the inherent sparsity of THz channels by learning to focus on significant propagation paths.
Simulation results demonstrate the superiority of our approach, as FP-ANet consistently outperforms several state-of-the-art methods across the entire tested SNR range. To ensure the reproducibility of our research, the code for the proposed model will be made publicly available upon acceptance of the paper.

\newpage
\bibliographystyle{IEEEbib}
\bibliography{strings,refs}

\end{document}